\documentclass[12pt]{iopart}

\usepackage{graphicx}
\usepackage{cite}
\usepackage[british]{babel}
\usepackage[T1]{fontenc}
\usepackage{xcolor}

\definecolor{red}{rgb}{0.0,0.0,0.0}
\definecolor{blue}{rgb}{0.0,0.0,0.0}

\begin{document}

\title[Heterogeneous diffusion in an harmonic potential: the role of the
interpretation]{Heterogeneous diffusion in an harmonic potential: the role
of the interpretation}

\author{Adrian Pacheco-Pozo$^1$, Igor M. Sokolov$^2$, Ralf Metzler$^3$, and
Diego Krapf$^1$}
\address{$^1$ Department of Electrical and Computer Engineering and School of Biomedical Engineering, Colorado State University, Fort Collins, CO 80523, USA.}
\address{$^2$ Institut f\"{u}r Physik and IRIS Adlershof, Humboldt University Berlin, Newtonstrasse 15, 12489 Berlin, Germany}
\address{$^3$ Institute of Physics and Astronomy, University of Potsdam, 14476 Potsdam, Germany and Asia Pacific Centre for Theoretical Physics, Pohang 37673, Republic of Korea}

\ead{ralf.metzler@uni-potsdam.de (corresponding author),
diego.krapf@colostate.edu}

\begin{abstract}
Diffusion in heterogeneous energy and diffusivity landscapes is widespread
in biological systems. However, solving the Langevin equation in such
environments introduces ambiguity due to the interpretation parameter
$\alpha$, which depends on the underlying physics and can take values in the
range $0<\alpha<1$. The typical interpretations are It\^o ($\alpha=0$),
Stratonovich ($\alpha=1/2$), and H\"anggi-Klimontovich ($\alpha=1$). Here,
we analyse the motion of a particle in an harmonic potential---modelled as an
Ornstein-Uhlenbeck process---with diffusivity that varies in space. Our focus
is on two-phase systems with a discontinuity in environmental properties
at $x=0$. We derive the probability density of the particle position for
the process, and consider two paradigmatic situations. In the first one, the
damping coefficient remains constant, and fluctuation-dissipation relations
are not satisfied. In the second one, these relations are enforced, leading
to a position-dependent damping coefficient. In both cases, we provide
solutions as a function of the interpretation parameter $\alpha$, with
particular attention to the It\^o, Stratonovich, and H\"anggi-Klimontovich
interpretations, revealing fundamentally different behaviours, in particular
with respect to an interface located at the potential minimum.
\end{abstract}

\section{Introduction}

Understanding the erratic motion of a particle in complex environments is a
problem that has attracted a great deal of attention over the last decade
\cite{Hanggi2005}. One of the typical tools used to study the diffusion
of particles is the Langevin equation, where the interaction between the
particle of interest and the environment is modelled via a random force 
\cite{Langevin1908,vanKampen1992,Ebeling2005}. The Langevin description \textcolor{blue}{can} naturally incorporate an external force acting on the tracer particle. \textcolor{blue}{A key distinction between the random and external forces lies in their timescales: the random force arises from rapid molecular collisions and fluctuates on very short timescales, while the external force is typically deterministic and varies more slowly.} Of particular interest is the case of a particle subjected to a restoring force,
as that produced by optical \cite{Ashkin1986,Keyser2006,Jeon2013,Heller2014}
and magnetic tweezers \cite{Smith1992,Vilfan2009,DeVlaminck2012}, and
tethered particle motion \cite{Schafer1991,Nelson2006,Dietrich2009}. The
stochastic process that describes the motion of a Brownian particle in
such a trapping potential is known as the Ornstein-Uhlenbeck process
\cite{Uhlenbeck1930,Doob1942,Wang1945}. This process is also widespread
beyond single-molecule biophysics and it is applied, for example, in finance
\cite{Barndorff2002,Maller2009}, evolution \cite{Bartoszek2017}, and chemical
kinetics \cite{Hanggi1990}.

\textcolor{blue}{The Langevin equation provides a framework for modelling the diffusion of particles in complex, heterogeneous media.} In
such environments, the strength of the random force may be time-
\cite{Chechkin2017,Sposini2024PRE,Sposini2024} and/or space-dependent
\cite{Fa2003,Farango2014,Uneyama2015,Leibovich2019,Pacheco2022,Pacheco2023,MenonJr2023,Pacheco2024,Li2024,Miler2024},
as opposed to being constant. The solution of the Langevin equation
involves a stochastic integral, which is well-defined if the
strength of the random force is constant or depends solely on time
\cite{vanKampen1981}. However, when the strength depends on the position,
the stochastic integral becomes ill-defined due to the extra freedom
in choosing how to evaluate the noise at each infinitesimal random increment
\cite{vanKampen1981,Tsekov1997,Sokolov2010,Sancho2011,deHaan2012,Arenas2012,Farango2014,dePirey2022,Escudero2023,Pacheco2024}.
To address this issue, a rule (interpretation) must be provided \textit{a
priori} on where to evaluate the noise strength. Due to their practical
usefulness, three particular interpretations are typically considered,
the initial-point It\^o interpretation \cite{Ito1944}, the mid-point
Stratonovich interpretation \cite{Stratonovich1964,Stratonovich1966},
and the final-point H\"anggi-Klimontovich interpretation
\cite{Hanggi1978,Hanggi1980,Hanggi1982,Klimontovich1994}. It is now accepted
\cite{vanKampen1981}  that there is no universal correct interpretation;
rather, the choice depends on the modelled system. For example, it was shown
experimentally that the H\"anggi-Klimontovich interpretation is the most
appropriate for modelling colloidal particles near a wall \cite{Volpe2010},
while the It\^o interpretation is necessary for modelling Brownian yet
non-Gaussian diffusion \cite{Postnikov2020}.

Although the question of a correct interpretation sometimes arises for
homogeneous systems (e.g., in a magnetic field \cite{Roma2014}), it is always
an issue for heterogeneous ones.  Here the disordered situations are the most
complex ones. Disordered systems may be classified as annealed or quenched.
Annealed disorder refers to temporal changes and a particle that returns to a
previously visited site does not necessarily experience the same interactions
\cite{Stauffer2005,Chechkin2017,balcerek2023modelling,Pacheco2024fBm}. In
contrast, quenched disorder refers to heterogeneities that are static in
time and depend solely on the position within the medium. Quenched disorder
is often more complicated to treat analytically since returns to already
visited locations introduce correlations in the particle motion. Hence,
numerical approximations are usually employed to study such systems. Among
the quenched disorder models, the quenched trap model has been employed broadly
\cite{Zarlenga2007,burov2007occupation,Krusemann2014,Miyaguchi2015,Akimoto2016,Akimoto2020,Shafir2024}.
This model is related to the continuous-time random walk
\cite{korabel2010paradoxes,burov2011time,Burov2017} and has proven to be
powerful from both theoretical and experimental perspectives.  To consider
the effects of a force in quenched systems, one usually use the Langevin
equation. However, in this case, the role of the interpretation remains
unresolved.

In order to gain insight into quenched heterogeneous systems, in this
article, we focus on a simple class of heterogeneity,  represented by a
position-dependent diffusion coefficient.  We study the effect of a linear
restoring force, i.e., an Ornstein-Uhlenbeck process, in a heterogeneous
environment by employing the Langevin equation. In this sense, we consider the
problem of the motion of a particle under the action of a restoring force in
a landscape characterised by a position-dependent diffusion coefficient. We
first consider the general problem for any given diffusion landscape as a
function of the interpretation parameter. Then, we focus on the case where the
diffusion coefficient is a piecewise-constant function, a simple heterogeneous
system, with a single interface. Despite its simplicity, this system yields
surprising results. We consider the interpretation as a parameter of our
model instead of choosing an \textit{a priori} interpretation. Particular
attention is paid to the It\^o, Stratonovich, and H\"anggi-Klimontovich
interpretations and the evolution of the system in time in these three
classical interpretations. Finally, we discuss conditions under which this
heterogeneous system obeys the fluctuation-dissipation relation with constant
temperature \cite{shemer2009einstein}.

\section{Homogeneous Ornstein-Uhlenbeck process}

We first consider the diffusion of a particle in a homogeneous environment subjected to a harmonic potential in which the force varies linearly with the distance from the origin, i.e., $F(x) = - k x$, with $k$ being the restoring force constant (Hooke coefficient). The equation describing the motion of a particle in such a system is the one-dimensional Langevin equation of the Ornstein-Uhlenbeck (OU) process \cite{Balakrishnan2008}, namely,
\begin{equation}
\dot{x} = - \frac{x}{\tau} + \sqrt{2D} \, \xi(t),
\label{eq:OU_SDE}
\end{equation}
where $\tau$ is the correlation time, $D$ is the diffusion coefficient, and $\xi(t)$ is a zero-mean Gaussian white noise with $\delta$-correlations. The correlation time is related to the constant $k$ via the damping coefficient $\gamma$ by $\tau =  \gamma/k$. The solution of this equation for the initial condition $x(0) = 0$ is given by
\begin{equation}
x(t) = \sqrt{2D} \int_0^t e^{- (t-t^{\prime})/\tau } \, \xi(t^{\prime}) \, dt^{\prime}.
\end{equation}
From this solution, it can be easily verified that the mean is zero ($\langle x(t) \rangle = 0$) and the covariance function takes the form
\begin{equation}
\langle x(t) x(t') \rangle = D \tau \left[ e^{- |t-t'|/\tau } - e^{- (t+t')/\tau } \right],
\end{equation}
from which the MSD is $\langle x^2(t) \rangle = D \tau \left[ 1 - e^{- 2t/\tau } \right]$. As the process approaches thermal equilibrium, the MSD converges to  $\langle x^2(t) \rangle = D \tau$. In the limit $\tau \to \infty$, i.e., a vanishing potential, the OU process reduces to standard Brownian motion, being the solution of the over-damped Langevin equation.

Associated to the Langevin equation of the OU process is a Fokker-Planck equation for the probability density function (PDF) $p(x,t)$ of the form 
\begin{equation}
\frac{\partial}{\partial t} p(x,t) = \frac{1}{\tau}\frac{\partial}{\partial x} [ x p(x,t)] + D \frac{\partial^2}{\partial x^2} p(x,t).
\label{eq:FP_HOU}
\end{equation}
The solution of this Fokker-Planck equation is \cite{Doi1988,Risken1996,Lidner2013}
\begin{equation}
p(x,t)=\frac{1}{\sqrt{2\pi D\tau S(t)}}\exp\left(-\frac{x^2}{\displaystyle 2D
\tau S(t)}\right),
\label{eq:OU_gen_PDF}
\end{equation}
where $S(t)= 1 - e^{-2t/\tau}$. In the limit $t \to \infty$, the system equilibrates and the steady state PDF takes the form
\begin{equation}
p(x) =  \frac{1}{\left( 2 \pi D \tau \right)^{1/2} } \exp \left( - \frac{x^2}{ 2 D \tau } \right),
\label{eq:OU_proc_stat}
\end{equation}
which is the Boltzmann distribution with
\begin{equation}
D\tau = \frac{k_B T}{k},
\label{eq:FD_r}
\end{equation}
\textcolor{blue}{where $T$ is the temperature and $k_B$ is the Boltzmann constant.
Eq.~(\ref{eq:FD_r}) is equivalent to the Einstein-Smoluchowski relation, a special
case of the fluctuation-dissipation theorem of the second kind, $D=\mu k_BT$ with
the mobility $\mu=(\tau k)^{-1}$.}

\section{Heterogeneous Ornstein-Uhlenbeck process \label{sec:tau_const}}

We now consider the case when the diffusion coefficient depends on the position,
i.e., $D(x)$. \textcolor{blue}{Analogously, in the context of annealed disorder,
a time-dependent diffusivity $D(t)$ was considered by Lanoisel\'ee et al. for
the motion of a particle in a harmonic potential (OU process)
\cite{Lanoiselee2022}.} The Langevin equation~(\ref{eq:OU_SDE}) for the
\textit{homogeneous} OU process needs to be modified, leading to the Langevin
equation for the \textit{heterogeneous} OU process
\begin{equation}
\dot{x}=-\frac{x}{\tau}+\sqrt{2D(x)}\,\xi(t).
\label{eq:OU_postion}
\end{equation}
While the diffusion coefficient possesses a spatial dependence, the
correlation time $\tau$ is kept constant. \textcolor{blue}{A constant
correlation time implies that either the fluctuation-dissipation relation
(Eq.~(\ref{eq:FD_r})) breaks down or that the effective temperature depends
on position. The former case is relevant to non-physical systems such as
fluctuations in financial data \cite{balcerek2025two}. In the latter case,
a position-dependent temperature $T(x)$ can reach a steady state in systems
away from equilibrium. For example, the local temperature increases when
a laser heats up a fluid, leading to significant temperature gradients and,
in turn, effects changes in the transport properties \cite{keyser2005nanopore}.
Another case in which a constant $\tau$ is relevant involves light-driven
active colloids. In these systems, the intensity and/or wavelength of the
incident light control the local stochastic driving force, and therefore
the diffusion coefficient, while not considerably influencing the mobility
\cite{Palacci2014,Vutukuri2020,Wang2024}. The constant-temperature situation
in which the fluctuation-dissipation relation holds and the correlation time
depends on position, $\tau(x)$, is explicitly discussed in Section 5.}

Given the position dependence of the diffusivity, the integral that solves
this stochastic differential equation is ill-defined due to the inherent
freedom in evaluating it. This ambiguity arises in Langevin equations with a
position-dependent noise term, also known as multiplicative noise, as opposed
to additive noise where it does not depend on the position. One thus needs
to provide a rule, or interpretation, to solve the above stochastic integral
\cite{vanKampen1981,Sokolov2010,Pacheco2024,dePirey2022}. \textcolor{blue}{In
this work, we treat the interpretation as a variable, maintaining generality,
and without focusing on any specific physical system, as doing so would
require selecting a particular interpretation \cite{Pacheco2024}.}

The interpretation explicitly appears in the corresponding Fokker-Planck
equation. In the heterogeneous OU process, the Fokker-Planck equation reads
\cite{Arenas2012,dePirey2022}
\color{blue}
\begin{equation}
\frac{\partial}{\partial t}p(x,t)=\frac{\partial}{\partial x}\left\{\frac{x}{
\tau}p(x,t)+D^\alpha(x)\frac{\partial}{\partial x}\left[D^{1-\alpha}(x)p(x,t)
\right]\right\},
\label{eq:FP_alpha}
\end{equation}
where $\alpha\in[0,1]$ is the interpretation parameter. For the three typical
interpretations mentioned in the introduction the interpretation parameters are
$\alpha=0$, $1/2$, and $1$ for It\^o, Stratonovich and H\"anggi-Klimontovich,
respectively \cite{Sokolov2010}. Moreover, in the case when both the
position-dependent diffusion coefficient and the PDF are differentiable
functions with respect to the position, then the Fokker-Planck reduces to
\begin{equation}
\frac{\partial}{\partial t}p(x,t)=\frac{\partial}{\partial x}\left[\frac{x}{
\tau}+(1-\alpha)\frac{\partial D(x)}{\partial x}+D(x)\frac{\partial}{\partial
x}\right]p(x,t),
\label{eq:FP_alpha_derivatives}
\end{equation}
where one can define an effective potential $V_{\rm eff}(x)$ so as to rewrite
it in the form \cite{Arenas2012,dePirey2022}
\color{black}
\begin{equation}
\frac{\partial}{\partial t} p(x,t) = \frac{\partial}{\partial x} \left[ \frac{\partial V_{\rm eff}(x)}{\partial x} + D(x) \frac{\partial}{\partial x} \right] p(x,t),
\end{equation}
where $\alpha$ modifies the effective potential,
\begin{equation}
V_{\rm eff}(x) = \frac{x^2}{2 \tau} + (1 - \alpha ) D(x).
\end{equation}
The contribution $(1 - \alpha ) D(x)$ is sometimes called the spurious drift. If the diffusion coefficient $D(x)$ is constant, the spurious drift is zero and the Fokker-Planck equation becomes independent of $\alpha$. Interestingly, when the interpretation is H\"anggi-Klimontovich ($\alpha = 1$), the effective potential reduces to the harmonic one. 

The stationary solution $p_{\rm st}(x)$ follows by taking the limit $t \to \infty$ in the above Fokker-Planck equation (Eq.~(\ref{eq:FP_alpha})), 
\color{blue}
\begin{equation}
\frac{d}{dx}\left\{\frac{x}{\tau}p_{\rm st}(x)+D^\alpha(x)\frac{d}{dx}\left[
D^{1-\alpha}(x)p_{\rm st}(x)\right]\right\}=0,
\end{equation}
from which, 
\begin{equation}
\frac{x}{\tau}p_{\rm st}(x)+D^\alpha(x)\frac{d}{dx}\left[D^{1-\alpha}(x)p_{\rm
st}(x)\right]=-\mathcal{J},
\end{equation}
\textcolor{red}{where $\mathcal{J}$ is a constant flux that must vanish in the
absence of sources and sinks.}
\color{black}
Then, the solution is \cite{dePirey2022}
\begin{equation}
p_{\rm st}(x)=\mathcal{N}D^{\alpha-1}(x)\exp\left[-\frac{1}{\tau}\int^x\frac{
x'}{D(x')}dx'\right],
\label{eq:PDF_eq_position}
\end{equation}
where $\int^x$ denotes the indefinite integral and $\mathcal{N}$ is a
normalisation constant.

\section{Ornstein-Uhlenbeck with piecewise-constant diffusion coefficient}

Let us now consider a piecewise constant diffusion coefficient of the form
\begin{equation}
\label{eq:piecewise}
D(x)=\left\{\begin{array}{ll}D_+&\mbox{for }x\geq0,\\\\
D_-&\mbox{for }x< 0,\end{array}\right.
\end{equation}
with $D_{\pm}>0$. \textcolor{blue}{Note that, due to the discontinuity at
$x=0$, the position-dependent diffusion coefficient is not differentiable
everywhere. As a result, we cannot directly use the Fokker-Planck
in the form of Eq.~(\ref{eq:FP_alpha_derivatives}), which assumes
differentiability. Instead, we must use the form of Eq.~(\ref{eq:FP_alpha}).}
This form of position-dependent diffusion coefficient was previously used
to find the solution of the Langevin equation in an open system, without
a trapping potential \cite{Pacheco2024}.  For such a diffusion landscape,
the steady-state PDF in Eq.~(\ref{eq:PDF_eq_position}) takes the form
\begin{equation}
\label{eq:stat_sol}
p_{\rm st}(x)=\left\{\begin{array}{lc}\displaystyle\frac{2\beta(\alpha)}{
\sqrt{2\pi D_-\tau}}\exp\left(-\frac{x^2}{2D_-\tau}\right),&\mbox{for }x<0,\\\\
\displaystyle\frac{2[1-\beta(\alpha)]}{\sqrt{2\pi D_+\tau}}\exp\left(-\frac{
x^2}{2D_+\tau}\right),&\mbox{for }x\geq0,\end{array}\right.
\end{equation}
where
\begin{equation}
\label{eq:beta}
\beta(\alpha)=\left[1+\left(\frac{D_-}{D_+}\right)^{(1/2-\alpha)}\right]^{-1}
\end{equation}
\textcolor{blue}{is the probability of finding a particle on the negative side
of the $x$-axis \cite{Pacheco2024}.  The stationary solution $p_{\rm st}(x)$}
is a generalised two-piece Gaussian distribution, similar to the time-dependent
PDF obtained without a trapping potential \cite{Pacheco2024}. The absence
of a trapping potential corresponds, in the OU process, to the limit $\tau
\to\infty$. In such a case, the time-dependent PDF is \cite{Pacheco2024}
\begin{equation}
\label{eq:BM_sol}
p_{\tau\to\infty}(x,t)=\left\{\begin{array}{lc}\displaystyle\frac{2\beta(
\alpha)}{\sqrt{4\pi D_-t}}\exp\left(-\frac{x^2}{4D_-t}\right),&\mbox{for }
x<0,\\\\\displaystyle\frac{2[1-\beta(\alpha)]}{\sqrt{4\pi D_+t}}\exp\left(
-\frac{x^2}{4D_+t}\right),&\mbox{for }x\geq0,\end{array}\right.
\end{equation}
with the same $\beta(\alpha)$ as in Eq.~(\ref{eq:beta}). The overall solution
for $t \ll \tau$ must have the form of Eq.~(\ref{eq:BM_sol}) because the
trapping potential does not play any role at short times, and for $t \gg
\tau$ must converge to the steady state solution (Eq.~(\ref{eq:stat_sol})).
Given the PDF of the homogeneous OU process, the solution of the Fokker-Planck
equation (Eq.~(\ref{eq:FP_alpha})) for the heterogeneous OU process with
piecewise diffusion coefficient for a particle starting at $x=0$ has the form
\begin{equation}
p_{x_0=0}(x,t)=\left\{\begin{array}{lc} 
\displaystyle\frac{2\beta(\alpha)}{\sqrt{2\pi D_-\tau S(t)}}\exp\left[- \frac{
x^2}{2D_-\tau S(t)}\right], & x < 0, \\\\
\displaystyle\frac{2[1-\beta(\alpha)]}{\sqrt{2\pi D_+\tau S(t)}}\exp\left[
-\frac{x^2}{2D_+\tau S(t)}\right], & x \geq 0.
\end{array} 
\right.
\label{eq:gen_PDF}
\end{equation}
Here, analogous to the homogeneous OU process, $S(t) = 1-e^{-2 t /
\tau}$. The situation with a different initial condition is discussed
below. The term $\beta(\alpha)$ is obtained from the continuity of the
flux (particles conservation) \cite{Pacheco2024} and has the same form as
in Eq.~(\ref{eq:beta}) (see~\textcolor{blue}{\ref{ap:flux}} for the full
proof). \textcolor{blue}{Indeed, since the probability $\beta(\alpha)$
of finding a particle on the negative $x$-axis does not depend on the
magnitude of the restoring force (i.e., it is independent of $\tau$), we
conclude that the presence of a restoring potential does not affect the
probability of finding a particle on the negative half-axis.} Furthermore,
for a particle that starts at the origin, this probability does not depend
on time. Later in the article, we discuss a particle starting away from the
origin, in which case this probability is not stationary.

\begin{figure}
\centering
\includegraphics[width=\textwidth]{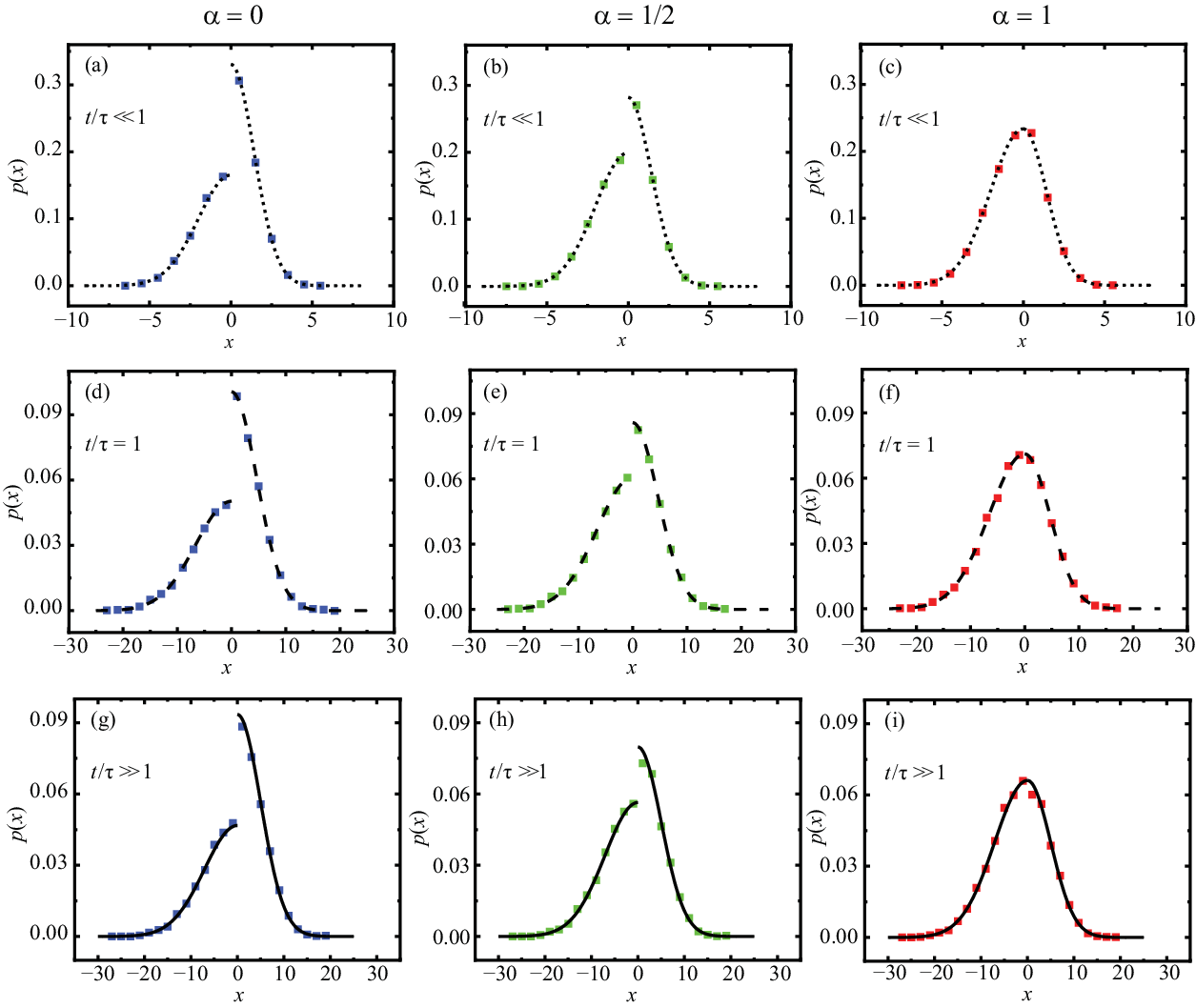}
\caption{PDFs of the heterogeneous OU process with $\tau=25$ and piecewise
diffusion coefficient with $D_-=2$ and $D_+=1$ for the interpretation
parameters $\alpha=0$, $1/2,$ and $1$, corresponding to the It{\^o},
Stratonovich, and H{\"a}nggi-Klimontovich interpretations, respectively. (a-c)
PDFs at time $t=1$ so that $t/\tau=0.04\ll1$. The dotted lines are
the PDFs of Brownian motion, Eq.~(\ref{eq:BM_sol}). (d-f) PDFs at time
$t=25$ so that $t/\tau=1$. The dashed lines are analytical solutions,
Eq.~(\ref{eq:gen_PDF}). (g-i) PDFs at time $t=1000$ so that $t/\tau=40\gg1$.
The solid lines are the steady state solutions, Eq.~(\ref{eq:stat_sol}).}
\label{fig:PDF}
\end{figure}

To validate our solution, we numerically simulate trajectories
according to the Langevin equation (Eq.~(\ref{eq:OU_postion})) for
the piecewise diffusion coefficient for three interpretations: It\^o,
Stratonovich, and H\"anggi-Klimontovich. We modified the Heun algorithm
\cite{Ruemelin1982,Garcia2012}, which implies the Stratonovich interpretation,
to solve the Langevin equations with any interpretation. This generalised
Heun algorithm is presented in~\textcolor{blue}{\ref{ap:mHeun}}. Using
this generalised algorithm, we simulate $10^4$ trajectories on a
diffusivity landscape with $D_-=2$ and $D_+=1$ with an integration
step of $\Delta t=10^{-2}$ and a correlation time $\tau=25$. These
trajectories are then used to construct the position PDF of the particles
at different times $t$. Figure~\ref{fig:PDF} shows the PDFs obtained
via numerical simulations together with the analytical solutions for
the It\^o, Stratonovich, and H\"anggi-Klimontovich interpretations
(Eq.~(\ref{eq:gen_PDF})). Figures~\ref{fig:PDF}(a-c) show the PDFs at time
$t=1$ such that $t/\tau=0.04\ll1$. The dotted lines in these figures
correspond to the PDFs of Brownian motion without a trapping potential
(Eq.~(\ref{eq:BM_sol})). Figures~\ref{fig:PDF}(d-f) present the PDFs
at time $t=25$, for which $t/\tau=1$. The dashed lines correspond
to the analytical solutions given by Eq.~(\ref{eq:gen_PDF}). Finally,
Figures~\ref{fig:PDF}(g-i) display the steady state PDFs at time $t=1000$
such that $t/\tau=40 \gg 1$. The solid lines correspond to the steady state
solutions of Eq.~(\ref{eq:stat_sol}).  Perfect agreement is observed between
the numerical simulations and the analytical solutions. \textcolor{blue}{The
discontinuity in the PDF for $\alpha<1$ is a consequence of the discontinuity
of $D(x)$. This observation is further addressed in~\ref{ap:continuity},
where we examine the effect of smoothing the step function with a continuous
approximation.}

From the PDF in Eq.~(\ref{eq:gen_PDF}), the mean and the MSD can be computed,
\begin{equation}
\langle x(t)\rangle_{x_0=0}=A(\alpha)\left[\tau(1-e^{-2t/\tau})\right]^{1/2},
\end{equation}
and 
\begin{equation}
\langle x^2(t)\rangle_{x_0=0}=B(\alpha)\,\tau(1-e^{-2t/\tau}),
\end{equation}
with 
\begin{equation}
A(\alpha)=\sqrt{\frac{2}{\pi}}\left[\sqrt{D_+}-\beta(\alpha)\left(\sqrt{D_+}
+\sqrt{D_-}\right)\right]
\end{equation}
and
\begin{equation}
B(\alpha)=D_++\beta(\alpha)\left(D_--D_+\right).
\end{equation}
For $t\to0$, the mean and MSD reduce respectively to $\langle x(t)
\rangle_{x_0=0}=A(\alpha)\sqrt{2t}$ and $\langle x^2(t)\rangle_{x_0=0}=2B(
\alpha)t$, which are the quantities obtained in the absence of
a trapping potential \cite{Pacheco2024}. \textcolor{blue}{In this case,
there is an initial flux in the direction of the larger diffusivity, except
when $\alpha=0$, in which case the flux vanishes.} On the other hand, for
$t\to\infty$, the mean and MSD reach a steady state with $\langle x(t)
\rangle_{x_0=0}=A(\alpha)\tau^{1/2}$ and $\langle x^2(t)\rangle_{x_0=0}
=B(\alpha)\tau$.

\begin{figure}
\centering
\includegraphics[width=\columnwidth]{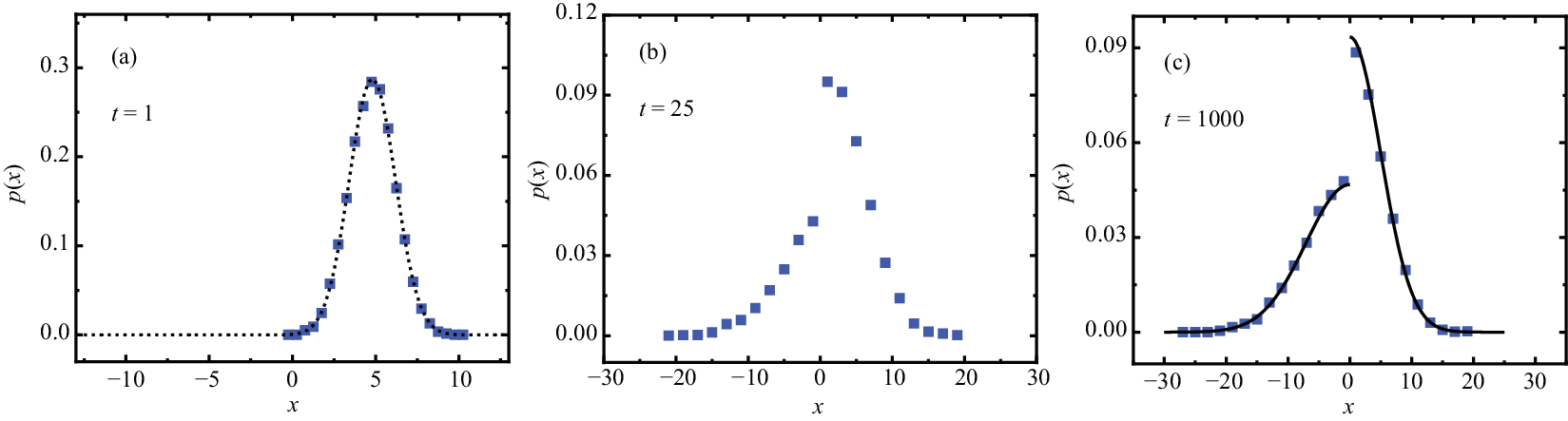}
\caption{PDFs of the heterogeneous OU process with a piecewise diffusion
coefficient with $D_-=2$ and $D_+=1$ for the It\^o interpretation
with initial condition $x_0=5$. In the simulations we set $\tau=25$.
\textcolor{blue}{PDF at times (a) $t=1$, (b) $t=25$, and (c) $t=1000$.
The dotted line is the PDF of the homogeneous OU process,
Eq.~(\ref{eq:free_OU}), with $D_+=1$ at $t=1$}. The solid line shows
the steady state PDF, Eq.~(\ref{eq:stat_sol}).}
\label{fig:PDF_initial}
\end{figure}

We now relax the assumption that the particles start at $x_0=0$, where the
position-dependent diffusion coefficient $D(x)$ presents a discontinuity and
the potential has its minimum. Then, for short times $t$ such that $t\ll\tau$
and $|x_0|\gg\sqrt{2D_\pm t}$, the PDF of the displacements is described by
\begin{equation}
p_{t\to0}(x,t)=\frac{1}{\sqrt{2\pi D_\pm\tau S(t)}}\exp\left(-\frac{(x-x_0
e^{-t/\tau})^2}{2D_\pm\tau S(t)}\right),
\label{eq:free_OU}
\end{equation}
for an initial condition $p(x,0) = \delta(x-x_0)$, $D_\pm=D_+$ when $x_0>0$,
and conversely $D_\pm = D_-$ for $x_0<0$. This PDF is essentially a Gaussian
profile centred at $x_0 e^{-t/\tau}$ and corresponds to the solution of the
OU process with a constant diffusion coefficient for an arbitrary initial
condition. In the short time limit, the probability of finding a particle
in the semi-infinite line opposite to the initial position is practically
zero and can be neglected, regardless of the interpretation.

As the system evolves, the centre of the Gaussian profile moves towards
$x = 0$, the potential minimum and where the diffusion coefficient has
its discontinuity. Due to this discontinuity, particles on either side
of $x = 0$ will move with different diffusion coefficients. As a result,
a generalised two-piece Gaussian distribution forms, where the probability
of being in the negative region depends on the interpretation. Furthermore,
since the initial condition is not at the origin, this probability varies
with time. In the long time limit, the memory of the initial condition is
lost, and the PDF takes the form of Eq.~(\ref{eq:stat_sol}).

\begin{figure}
\centering
\includegraphics[width=0.5\columnwidth]{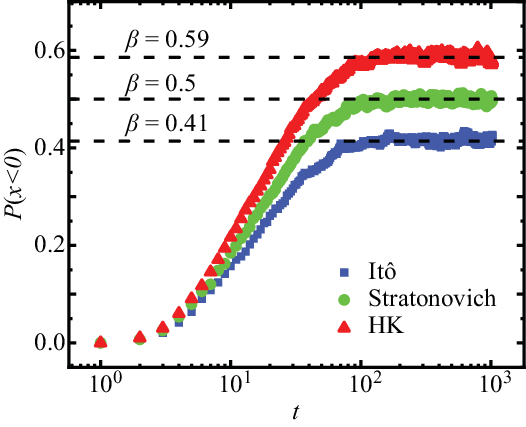}
\caption{Temporal evolution of the probability of finding a particle in
the negative part of the $x$-axis, $P(x<0)$, for the heterogeneous OU
process with a piecewise constant diffusion coefficient with $D_-=2$
and $D_+=1$ with It\^o, Stratonovich, and H\"anggi-Klimontovich (HK)
interpretations with initial condition $x_0=5$, and $\tau=25$. The
dashed lines corresponds to $\beta(\alpha)$ for $x_0=0$ (Eq.~(\ref{eq:beta}))
for the three interpretations.}
\label{fig:Prob}
\end{figure}

Finding a general solution of the Fokker-Planck equation
(Eq.~(\ref{eq:FP_alpha})) for an arbitrary non-zero initial condition is beyond
our abilities at the moment. However, we can perform numerical simulations
to gain insight into the behaviour of the PDF. Figure~\ref{fig:PDF_initial}
shows the evolution of the PDF for a system with $D_+=1$ and $D_-=2$
with initial condition $x_0=5$ under the It\^o interpretation. The
behaviour we describe here is qualitatively similar to the other
interpretations. In these simulations, we use the same parameters as
above. Figure~\ref{fig:PDF_initial}(a) shows the PDF at time $t=1$. In this
short time behaviour ($t\to0$), the particle is found around the starting
point and the probability of being in the negative part is essentially zero,
i.e., $P(x<0;t\to0)=0$. \textcolor{blue}{Figure~\ref{fig:PDF_initial}(b)
presents the PDF at time $t=25$. During this intermediate time, the
probability of being in the negative part of the $x$-axis grows with
time. Finally, figure~\ref{fig:PDF_initial}(c) depicts the PDF at time
$t=1000$.} In this long time limit ($t\to\infty$), one sees that the
initial condition is forgotten, and the PDF takes the asymptotic form of
Eq.~(\ref{eq:stat_sol}) with $\beta(\alpha)$ defined by Eq.~(\ref{eq:beta})
for $\alpha = 0$. In this asymptotic regime, the probability $P(x<0)$ reaches
a steady state equal to $\beta(\alpha)$. The evolution of the probability
of finding a particle in the negative part of the $x$-axis is presented
in Figure~\ref{fig:Prob}.

\section{Heterogeneous Ornstein-Uhlenbeck process with constant temperature \label{sec:T_const}}

\textcolor{blue}{Let us now consider the case of constant temperature,
for which the correlation time $\tau(x)$ is position dependent (cf.
section~\ref{sec:tau_const}) and} is given by
\begin{equation}
\label{eq:FD_rel}
\tau(x)=k_BT/kD(x),
\end{equation}
and the Langevin equation takes the form
\begin{equation}
\label{eq:LA_case2}
\dot{x}=-\frac{kxD(x)}{k_BT}+\sqrt{2D(x)}\xi(t).
\end{equation}
The associated Fokker-Planck equation is then
\color{blue}
\begin{equation}
\label{eq:FP_case2}
\frac{\partial}{\partial t}p(x,t)=\frac{\partial}{\partial x}\left(\frac{kxD(x)
}{k_BT}p(x,t)+D(x)\frac{\partial}{\partial x}\left[D^{1-\alpha}(x)p(x,t)\right]
\right).
\end{equation}
\color{black}
Following a similar procedure as we did before, we find the steady state
solution $p_{\rm st}^{\rm T}(x)$ in the limit $t\to\infty$,
\begin{equation}
p^{\rm T}_{\rm st}(x)=\mathcal{M}D^{\alpha -1}(x)\exp\left(-\frac{kx^2}{2k_BT}
\right),
\label{eq:PDF_eq_position_case_2}
\end{equation}
with the normalisation constant $\mathcal{M}$. The superscript $\rm{T}$ stands
for fulfilment of the fluctuation-dissipation relation (\ref{eq:FD_rel}) with
constant temperature.

\begin{figure*}
\centering
\includegraphics[width=\textwidth]{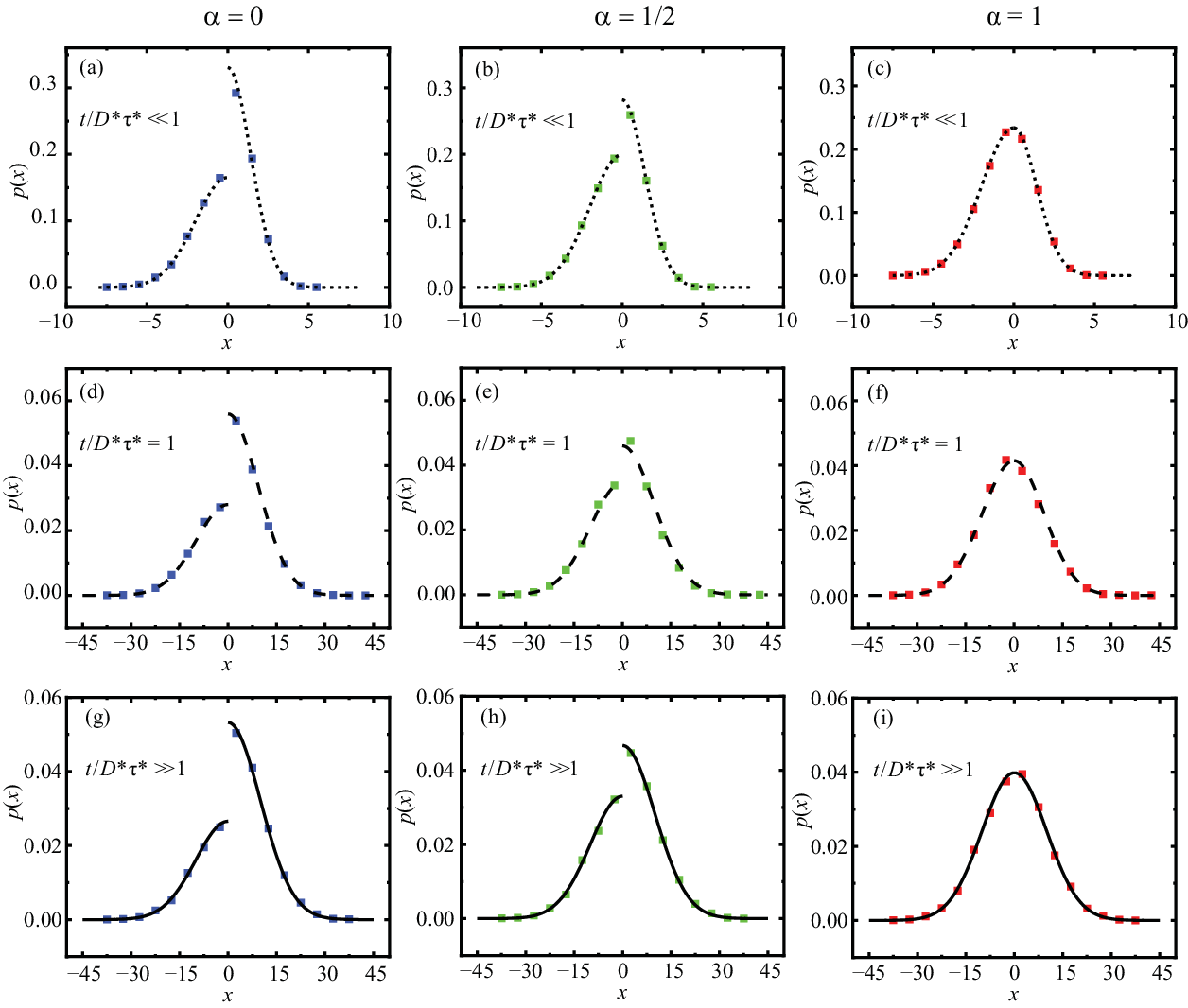}
\caption{PDFs of the heterogeneous OU process with a piecewise diffusion
coefficient, constant temperature, and fluctuation-dissipation
relations. The diffusivities are $D_-=2$ and $D_+=1$. In the simulations,
we set $D^*\tau^*=k_BT/k=100$. (a-c) PDFs at time $t=1$ so that $t/D^*\tau^*
=0.01\ll1$. The dotted lines are the PDF (\ref{eq:BM_sol}) of Brownian motion.
(d-f) PDFs at time $t=100$ so that $t/D^*\tau^*=1$. The dashed lines are the
analytical solutions, Eq.~(\ref{eq:gen_PDF_2}). (g-i) PDFs at time $t=1000$
so that $t/D^*\tau^*=10>1$. The solid lines represent the steady state PDF
(\ref{eq:PDF_HK_2}).}
\label{fig:PDF_2}
\end{figure*}

Let us consider again the piecewise-constant diffusion coefficient of
Eq.~(\ref{eq:piecewise}). For such a diffusion landscape, the steady state
PDF (\ref{eq:PDF_eq_position_case_2}) takes the form
\begin{equation}
\label{eq:case_2_ss}
p^{\rm{T}}_{\rm st}(x)=\left\{\begin{array}{lc}\displaystyle\frac{2\eta(\alpha)
}{\displaystyle\sqrt{2\pi k_BT/k}}\exp\left(-\frac{kx^2}{2k_BT}\right),&\mbox{
for }x<0,\\\\\displaystyle\frac{2[1-\eta(\alpha)]}{\displaystyle\sqrt{2\pi k_BT
/k}}\exp\left(-\frac{kx^2}{2k_BT}\right),&\mbox{for }x\geq0,\end{array}\right.
\end{equation}
with
\begin{equation}
\eta(\alpha) = \left[ 1 + \left( \frac{D_-}{D_+} \right)^{(1-\alpha)} \right]^{-1}.
\end{equation}
Note the different $\alpha$-dependence of $\eta(\alpha)$ when compared
to $\beta(\alpha)$ in Eq.~(\ref{eq:beta}). Despite the presence of the
heterogeneous environment, the PDF $p^{\rm{T}}_{\rm st}(x)$ is Gaussian in
the H\"anggi-Klimontovich interpretation ($\alpha=1$),
\begin{equation}
\label{eq:PDF_HK_2}
p^{\rm{T}}_{\rm st}(x)=\sqrt{\frac{k}{2\pi k_BT}}\exp\left(-\frac{kx^2}{2k_BT}
\right),
\end{equation}
which is the Boltzmann distribution, and not a two-piece Gaussian distribution
as in the case discussed above. It is the same PDF one obtains for the steady
state solution of the homogeneous OU process (\ref{eq:OU_proc_stat}).

We now find a solution for any $\alpha$ and any time with $x_0=0$. Let us
define $D^*\tau^*=k_BT/k$. Following a similar line of thought as
before, the overall solution must have the form of Eq.~(\ref{eq:BM_sol})
for $t\ll D^*\tau^*$, because the trapping potential does not play any
role at short times. It must also converge to the steady state solution
(Eq.~(\ref{eq:case_2_ss}) for $t\gg D^*\tau^*$. The solution of the
Fokker-Planck equation~(\ref{eq:FP_case2}) with fluctuation-dissipation,
for a piecewise constant diffusion coefficient for a particle starting at
$x_0=0$ is given by
\begin{equation}
p^{\rm{T}}_{x_0=0}(x,t)= \left\{ 
\begin{array}{lc} 
\displaystyle\frac{2\eta(\alpha,t)}{\sqrt{2\pi D^*\tau^* S_-(t)}}\exp\left[
-\frac{x^2}{2D^*\tau^*S_-(t)}\right], & x < 0, \\\\
\displaystyle\frac{2[1-\eta(\alpha,t)]}{\sqrt{2\pi D^*\tau^*S_+(t)}}\exp\left[
-\frac{x^2}{2D^*\tau^*S_+(t)}\right], & x \geq 0,
\end{array} 
\right.
\label{eq:gen_PDF_2}
\end{equation}
with
\begin{equation}
\eta(\alpha,t)=\left\{1+\left(\frac{D_-}{D_+}\right)^{1-\alpha}\left[\frac{S_
-(t)}{S_+(t)}\right]^{-1/2}\right\}^{-1},
\label{eq:gamma}
\end{equation}
and $S_{\pm}(t)=1-\exp(-2D_{\pm}t/D^*\tau^*)$. This form of $\eta(\alpha,t)$
is obtained, as for the previous case, by the continuity of the flux (see
Eq.~(\ref{eq:ap_eta}) in~\textcolor{blue}{\ref{ap:flux}}). Note that in this
case the probability of being in the negative part ($x<0$) depends on both
time and the interpretation $\alpha$, as opposed to the previous case in which
such a probability depends only on the interpretation. The mean and second
moment (MSD) of this process are presented in \textcolor{blue}{\ref{ap:mean}}.

To validate our solution, we numerically solved Eq.~(\ref{eq:LA_case2}) for
the piecewise diffusion coefficient using the modified Heun algorithm
(see~\textcolor{blue}{\ref{ap:mHeun}}) to obtain trajectories of
the process $x(t)$ in three interpretations: It\^o, Stratonovich and
H\"anggi-Klimontovich. In the simulations we set $D^*\tau^*=100$
and $\Delta t=5\times10^{-3}$. We simulate $10^4$ trajectories on
a diffusivity landscape with $D_-=2$ and $D_+=1$. Using these
trajectories, we construct the PDF with the position of the particles
at a given time $t$. Figure~\ref{fig:PDF_2} shows the PDFs obtained via
numerical simulations at different times along with the corresponding
analytical solutions. Figures~\ref{fig:PDF_2}(a-c) show the PDFs at time $t
=1$ such that $t/D^*\tau^*=0.01 \ll 1$. The dotted lines are the PDFs
of Brownian motion, Eq.~(\ref{eq:BM_sol}). Figures~\ref{fig:PDF_2}(d-f)
present the PDFs at time $t=100$, for which $t/D^*\tau^*=1$. The dashed
lines are the analytical solution, Eq.~(\ref{eq:gen_PDF_2}). Finally,
figures~\ref{fig:PDF_2}(g-i) display the PDFs at time $t=1000$ such that
$t/D^*\tau*=10\gg1$. The solid lines are the PDF of the steady state
solution, Eq.~(\ref{eq:PDF_HK_2}). Excellent agreement is again observed.

\section{Discussion and Conclusions}

In this work, we studied the motion of a particle in a harmonic potential
within a heterogeneous environment modelled by a position-dependent
diffusion coefficient. To account for the heterogeneous landscape, we
modified the Langevin equation of the homogeneous OU process so that the
diffusion coefficient depends on position, i.e., $D(x)$. We first consider
the problem for a general landscape $D(x)$ with constant correlation time
$\tau$, i.e., constant damping coefficient. Then, within the assumption
of a constant correlation time, we turned our attention to the case of a
piecewise constant diffusion coefficient (Eq.~(\ref{eq:piecewise})) with
a discontinuity at $x=0$, which is also the point of the minimum of the
potential. We first considered the situation where the particle's initial
position coincides with this discontinuity, i.e., $x_0=0$. This situation
is particularly interesting since it allows us to find an exact solution to
the Fokker-Planck equation. Remarkably, the obtained PDF has a similar form
as the PDF obtained without a harmonic potential \cite{Pacheco2024}, namely,
a generalised two-piece Gaussian distribution. Moreover, while in the presence
of the potential, the probability of finding a particle in the negative part
($x<0$) depends on $\alpha$, it does not change with time. This probability
is also not affected by the potential \cite{Pacheco2024}. We corroborated our
analytical solutions by numerically solving the corresponding Langevin equation
for the It\^o, Stratonovich, and H\"anggi-Klimontovich interpretations.

We further considered the solution where the initial condition is not at the
origin. We performed numerical simulations to obtain the PDF. It was found
that at short times, the PDF is described by the homogeneous OU process. On
the other hand, at large times, the PDF shows the same asymptotic behaviour
as in the case with the initial condition at $x_0=0$.

In the second part of this work, we consider the scenario in which the
fluctuation-dissipation relation is fulfilled under constant temperature,
which is the case for most single-molecule experiments. Thus, we analysed
the case where the damping coefficient depends on the position and is
inversely proportional to the diffusion coefficient. We found the exact
PDF that solves the Fokker-Planck equation for an initial condition at the
origin, $x_0=0$. Similar to the previous situation, the time-dependent
solution for this case also has the form of a generalised two-piece Gaussian
distribution. However, the probability of being in the negative part of
the $x$-axis is not constant but a time-dependent quantity that depends
on the interpretation parameter. Interestingly, the PDF convergences to a
normal distribution for the H\"anggi-Klimontovich interpretation ($\alpha =
1$). Again, we corroborated our solutions with numerical simulations of the
Langevin equation.

This work presents the analysis of the heterogeneous OU process where the
interpretation is considered as a parameter throughout the analysis. Finally,
this study paves the way for a possible generalisation to any potential,
or even, for cases where the system presents correlations, like in the case
of fractional Brownian motion.

\ack
This work was supported by the National Science Foundation (NSF) Grant 2102832
(to DK) as well as NSF-BMBF CRCNS (grant 2112862/STAXS) and German Science
Foundation (DFG, grants ME 1535/13-1 and ME 1535/22-1) (to RM).

\appendix

\section{Probability of being in $x<0$ \label{ap:flux}}

Let us start from the Fokker-Planck equation (\ref{eq:FP_alpha}) which we
rewrite in the form
\begin{equation}
\frac{\partial}{\partial t}p(x,t)=\frac{\partial}{\partial x}\left[\frac{x}{
\tau}p(x,t)\right]+\frac{\partial}{\partial x}\left\{D^\alpha(x)\frac{\partial
}{\partial x}\left[D^{1-\alpha}(x)p(x,t)\right]\right\},
\end{equation}
which is essentially a continuity equation and can be rewritten as
\begin{equation}
\frac{\partial}{\partial t}p(x,t)=-\frac{\partial}{\partial x}\mathcal{J}(x,t),
\end{equation}
with $\mathcal{J}(x,t)$ being the diffusive flux defined as
\begin{equation}
\mathcal{J}(x,t)=-\frac{x}{\tau}p(x,t)-D^\alpha(x)\frac{\partial}{\partial x}
\left[D^{1-\alpha}(x)p(x,t)\right].
\label{eq:flux}
\end{equation}
This is essentially a generalised form of Fick's first law that includes the
potential.

\textcolor{blue}{Given that the transport equation (in our case, the
Fokker-Planck equation) appears as a combination of a generalised continuity
equation (possibly, with sources), and a constitutive equation for the
flux (Eq.~(\ref{eq:flux})), the discontinuity of the flux would imply the
existence of a point source at the point of discontinuity. In our setup
the sources are, however, absent. Therefore, the flux is continuous and
differentiable with respect to $x$. Moreover, the definition of the flux
(Eq.~(\ref{eq:flux})) implies that the product $D^{1-\alpha}(x)p(x,t)$ must
itself be differentiable with respect to position. With these considerations
in place, we proceed to determine the probability $\beta(\alpha)$ of finding a
particle on the negative side of the $x$-axis. Since the diffusion coefficient
$D(x)$ is strictly positive, we can divide both sides of Eq.~(\ref{eq:flux})
by $D^\alpha(x)$, which yields}
\begin{equation}
-\frac{\mathcal{J}(x,t)}{D^\alpha(x)}=\frac{x}{\tau}\frac{p(x,t)}{D^\alpha(x)}
+\frac{\partial}{\partial x}\left[D^{1-\alpha}(x)p(x,t)\right].
\end{equation}
Next, we integrate both sides of this expression in a vicinity of the origin,
from $-\epsilon$ to $\epsilon$, and obtain
\begin{equation}
-\int_{-\epsilon}^{\epsilon}\frac{\mathcal{J}(x,t)}{D^\alpha(x)}dx=\int_{-
\epsilon}^{\epsilon}\frac{x}{\tau}\frac{p(x,t)}{D^\alpha(x)}dx+\int_{-
\epsilon}^{\epsilon}\frac{\partial}{\partial x}\left[D^{1-\alpha}(x)p(x,t)
\right] dx.
\end{equation}
\color{blue}
The last integral on the right-hand side can be readily computed since $D^{1-
\alpha}(x)p(x,t)$ is differentiable, therefore
\begin{equation}
\fl-\int_{-\epsilon}^{\epsilon}\frac{\mathcal{J}(x,t)}{D^\alpha(x)}dx=\int_{
-\epsilon}^{\epsilon}\frac{x}{\tau}\frac{p(x,t)}{D^\alpha(x)}dx+\left[ D^{1-
\alpha}(\epsilon) p(\epsilon,t)-D^{1-\alpha}(-\epsilon)p(-\epsilon,t)\right].
\label{eq:integrals}
\end{equation}
The remaining two integrals can be solved by means of the mean-value theorem:
Let $f(x)$ be a continuous function on $[a,b]$, and let $g(x)$ be integrable
and non-negative on $[a,b]$. Then, there exist some $c \in [a,b]$ such that
\begin{equation}
\int_a^bf(x)g(x)dx=f(c)\int_a^bg(x)dx.
\end{equation}
We can apply the mean-value theorem to the integrals in Eq.~(\ref{eq:integrals})
since both $\mathcal{J}(x,t)$ and $x$ are continuous functions, and $1/D^{\alpha}
(x)$ and $p(x,t)/D^{\alpha}(x)$ are integrable, non-negative functions. Thus,
for some $x'$ and $x''\in(-\epsilon,\epsilon)$, we have  
\color{black}
\begin{equation}
\fl-\mathcal{J}(x',t)\left(\frac{\epsilon}{D^\alpha_+}+\frac{\epsilon}{D^\alpha
_-}\right)=\frac{x''}{\tau}\int_{-\epsilon}^{\epsilon}\frac{p(x,t)}{D^\alpha(x)}
dx+\left[D^{1-\alpha}(\epsilon)p(\epsilon,t)-D^{1-\alpha}(-\epsilon)p(-\epsilon
,t)\right].
\end{equation}
Then, in the limit $\epsilon\to0$, the l.h.s.~vanishes. In this limit, the
first term on the r.h.s.~also vanishes since $x''\to0$. Hence,
\begin{equation}
D^{1-\alpha}_+p(0_+,t)=D^{1-\alpha}_-p(0_-,t).
\label{eq:condi}
\end{equation}
This shows that there is a discontinuity in the PDF for any $\alpha\neq1$. The
same expression holds if one starts from the Fokker-Planck equation
(\ref{eq:FP_case2}). However, in this case the PDF is different. Let us then
treat these two cases separately:

(i) Let us first consider the PDF in Eq.~(\ref{eq:gen_PDF}). This PDF at
$x=0_{\pm}$ reads
\begin{equation}
p(0_-,t)=\frac{2\beta(\alpha)}{\sqrt{2\pi D_-\tau S(t)}}
\end{equation}
and
\begin{equation}
p(0_+,t)=\frac{2[1-\beta(\alpha)]}{\sqrt{2\pi D_-\tau S(t)}}.
\end{equation}
Then, substituting these two expressions into Eq.~(\ref{eq:condi}), we obtain
\begin{equation}
\beta(\alpha)=\left[1+\left(\frac{D_-}{D_+}\right)^{1/2-\alpha}\right]^{-1}.
\end{equation}

(ii) Next, let us consider the PDF in Eq.~(\ref{eq:gen_PDF_2}). This PDF at
$x=0_{\pm}$ has the form 
\begin{equation}
p(0_-,t)=\frac{2\beta(\alpha)}{\sqrt{2\pi D^*\tau^*S_-(t)}}
\end{equation}
and
\begin{equation}
p(0_+,t)=\frac{2[1-\beta(\alpha)]}{\sqrt{2\pi D^*\tau^*S_+(t)}}.
\end{equation}
Next, we substitute these two expression into Eq.~(\ref{eq:condi}) and obtain
\begin{equation}
\eta(\alpha,t)=\left[1+\left(\frac{D_-}{D_+}\right)^{1-\alpha}\left(\frac{S_
-(t)}{S_+(t)}\right)^{-1/2}\right]^{-1}.
\label{eq:ap_eta}
\end{equation}

\section{Generalised Heun algorithm \label{ap:mHeun}}

The Heun algorithm \cite{Ruemelin1982,Garcia2012} is a
\textit{predictor-corrector} method that is used to numerically integrate a
stochastic differential equation (SDE) in the Stratonovich interpretation. The
Heun algorithm for a general SDE of the form
\begin{equation}
dX(t)=f[X(t),t]dt+g[X(t),t]dB(t),
\end{equation}
is a two-step process: First, one \textit{predicts\/} the next value using an
Euler scheme \cite{Garcia2012},
\begin{equation}
\overline{X}_{i+1}=X_i+f(X_i,t_i)\Delta t+g(X_i,t_i)\xi_i,
\label{eq:euler}
\end{equation}
where $X_i=X(t_i)$, and $t_{i+1}=t_i+\Delta t$. Then, one \textit{corrects\/}
this value for the Stratonovich interpretation using
\begin{eqnarray}
\nonumber
X_{i+1}&=&X_i+\frac{1}{2}\left[f(X_i,t_i)+f\left(\overline{X}_{i+1},t_{i+1}
\right)\right]\Delta t\\
&&+\frac{1}{2}\left[g(X_i,t_i)+g\left(\overline{X}_{i+1},t_{i+1}\right)\right]
\xi_i.
\end{eqnarray}
The proposed modified Heun algorithm is again a two part process: First,
using the same Euler scheme in Eq.~(\ref{eq:euler}), we predict the value
$\overline{X}_{i+1}$ after one time-step $t_{i+1}=t_i+\Delta t$. Then,
we modify the correction part of the standard Heun method to account for
the different interpretations as follows
\begin{eqnarray}
\nonumber
X_{i+1}&=&X_i+\left\{f(X_i,t_i)+\alpha\left[f\left(\overline{X}_{i+1},t_{i+1}
\right)-f(X_i,t_i)\right]\right\}\Delta t\\
&&+\left\{g(X_i, t_i)+\alpha\left[g\left(\overline{X}_{i+1},t_{i+1}\right)-
g(X_i,t_i)\right]\right\}\xi_i.
\end{eqnarray}
For the It\^o interpretation with $\alpha=0$, the correction step is
redundant and thus our modified Heun algorithm reduces to the standard Euler
scheme. For the Stratonovich case with $\alpha=1/2$, our modified scheme
reduces to the standard Heun algorithm. For the H\"anggi-Klimontovich with
$\alpha=1$, the new algorithm takes the form
\begin{equation}
X_{i+1}=X_i+f\left(\overline{X}_{i+1},t_{i+1}\right)\Delta t+ g\left(\overline{
X}_{i+1},t_{i+1}\right)\xi_i.
\end{equation}

\color{blue}

\section{Continuity of the PDF for smooth diffusion coefficient \label{ap:continuity}}

\begin{figure*}
\centering
\includegraphics[width=\textwidth]{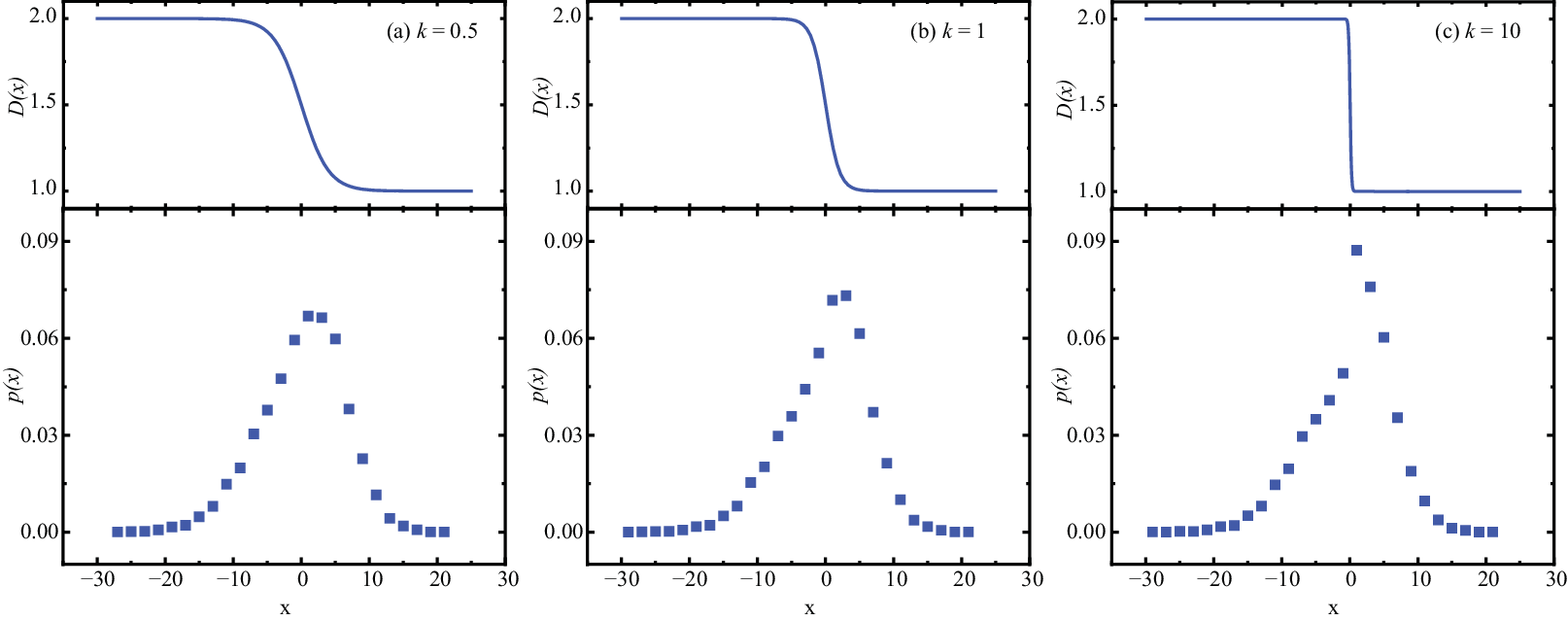}
\caption{PDFs of the heterogeneous OU process with continuous diffusion
coefficient defined in Eq.~(\ref{eq:cont_D}), shown for three values of
the parameter $k$: $0.5$, $1$, and $10$ corresponding to panels (a), (b), and
(c), respectively. Each panel also displays the corresponding diffusion
coefficient used in the simulations.}
\label{fig:continuous}
\end{figure*}

To better understand the origin of the discontinuity in $p(x,t)$, we examine
whether it arises from the discontinuity in the space-dependent diffusion
coefficient $D(x)$. Although this analysis can be carried out for both
versions of the heterogeneous OU process, we consider only the first case
for simplicity. We begin by rewriting the piecewise-constant diffusion
coefficient in Eq.~(\ref{eq:piecewise}) as
\begin{equation}
D(x)=(D_+-D_-)U(x)+D_-,
\end{equation}
where $U(x)$ is the Heaviside step function. Then, we consider a smooth
approximation of the Heaviside function in the form of a logistic function
\begin{equation}
\mathcal{U}(x)=\frac{1}{1+\exp(-kx)},
\end{equation}
which converges to $U(x)$ as $k\to\infty$. We then define the continuous
diffusion coefficient as
\begin{eqnarray}
D(x)=(D_+-D_-)\mathcal{U}(x)+D_-.
\label{eq:cont_D}
\end{eqnarray}

Using the generalised Heun algorithm, we computed $10^4$ trajectories for
three values of $k=0.5$, $1$, and $10$ under the It\^o interpretation, as
this is the case in which the PDF shows the greatest discontinuity for the
piecewise-constant diffusion coefficient (see Fig.~\ref{fig:PDF}). From these
trajectories, we calculate the PDF at time $t=1000$ for each value of $k$
and display it in figure~\ref{fig:continuous}, along with corresponding
continuous diffusion coefficient used in the simulations. As shown in
the figure, the PDF remains continuous when the diffusion coefficient is
continuous (see figures~\ref{fig:continuous}(a) and (b) for $k=0$, $5$,
and $1$, respectively). However, when the smooth diffusion coefficient
begins to closely resemble the discontinuous, as is already the case at
$k=10$, the PDF also becomes visibly discontinuous, as illustrated in
figure~\ref{fig:continuous}(c).

\color{black}

\section{Mean and MSD of the PDF in Eq.~(\ref{eq:gen_PDF_2}) \label{ap:mean}}

From the PDF in Eq.~(\ref{eq:gen_PDF_2}), the first two moments can be computed.
Then, the mean is
\begin{equation}
\fl\langle x(t)\rangle=\sqrt{\frac{2D^*\tau^*}{\pi}}\,\frac{D_-^{1-\alpha}\left(
1-e^{-2D_+t/D^*\tau^*}\right)-D_+^{1-\alpha}\left(1-e^{-2D_-t/D^*\tau^*}\right)}{
D_+^{1-\alpha}\left(1-e^{-2D_-t/D^*\tau^*}\right)^{1/2}+D_-^{1-\alpha}\left(1-e^{
-2D_+t/D^*\tau^*}\right)^{1/2}},
\end{equation}
and the MSD is
\begin{equation}
\fl\langle x^2(t)\rangle=D^*\tau^*\frac{D_-^{1-\alpha}\left(1-e^{-2D_+t/D^*\tau
^*}\right)^{3/2}+D_+^{1-\alpha}\left(1-e^{-2D_-t/D^*\tau^*}\right)^{3/2}}{D_+^{
1-\alpha}\left(1-e^{-2D_-t/D^*\tau^*}\right)^{1/2}+D_-^{1-\alpha}\left(1-e^{-2D_
+t/D^*\tau^*}\right)^{1/2}}.
\end{equation}

\section*{References}

\bibliographystyle{iopart-num.bst}
\bibliography{biblio}

\providecommand{\noopsort}[1]{}\providecommand{\singleletter}[1]{#1}%
\providecommand{\newblock}{}
\begin{thebibliography}{10}
\expandafter\ifx\csname url\endcsname\relax
  \def\url#1{{\tt #1}}\fi
\expandafter\ifx\csname urlprefix\endcsname\relax\def\urlprefix{URL }\fi
\providecommand{\href}[2]{#1}  
\providecommand{\eprint}[2][arXiv]{#1:\linebreak[0]#2}

\bibitem{Hanggi2005}
Hänggi P and Marchesoni F 2005 {\em Chaos\/} {\bf 15} 026101 ISSN 1054-1500

\bibitem{Langevin1908}
Langevin P 1908 {\em C. R. Acad. Sci. (Paris)\/} {\bf 146} 530--533

\bibitem{vanKampen1992}
Van~Kampen N~G 1992 {\em Stochastic processes in physics and chemistry\/} vol~1
  (Elsevier)

\bibitem{Ebeling2005}
Ebeling W and Sokolov I~M 2005 {\em Statistical Thermodynamics And Stochastic
  Theory Of Nonequilibrium Systems\/} Series On Advances In Statistical
  Mechanics (World Scientific Publishing Company) ISBN 9789813104631

\bibitem{Ashkin1986}
Ashkin A, Dziedzic J~M, Bjorkholm J~E and Chu S 1986 {\em Opt. Lett.\/} {\bf
  11} 288--290

\bibitem{Keyser2006}
Keyser U~F, Koeleman B~N, van Dorp S, Krapf D, Smeets R~M~M, Lemay S~G, Dekker
  N~H and Dekker C 2006 {\em Nat. Phys.\/} {\bf 2} 473--477 ISSN 1745-2481

\bibitem{Jeon2013}
Jeon J~H, Leijnse N, Oddershede L~B and Metzler R 2013 {\em New J. Phys.\/}
  {\bf 15} 045011

\bibitem{Heller2014}
Heller I, Hoekstra T~P, King G~A, Peterman E~J~G and Wuite G~J~L 2014 {\em
  Chem. Rev.\/} {\bf 114} 3087--3119 ISSN 0009-2665

\bibitem{Smith1992}
Smith S~B, Finzi L and Bustamante C 1992 {\em Science\/} {\bf 258} 1122--1126

\bibitem{Vilfan2009}
Vilfan I~D, Lipfert J, Koster D~A, Lemay S~G and Dekker N~H 2009 {\em Magnetic
  Tweezers for Single-Molecule Experiments\/} (New York, NY: Springer US) pp
  371--395 ISBN 978-0-387-76497-9

\bibitem{DeVlaminck2012}
De~Vlaminck I and Dekker C 2012 {\em Ann. Rev. Biophys.\/} {\bf 41} 453--472
  ISSN 1936-1238

\bibitem{Schafer1991}
Schafer D~A, Gelles J, Sheetz M~P and Landick R 1991 {\em Nature\/} {\bf 352}
  444--448 ISSN 1476-4687

\bibitem{Nelson2006}
Nelson P~C, Zurla C, Brogioli D, Beausang J~F, Finzi L and Dunlap D 2006 {\em
  J. Phys. Chem. B\/} {\bf 110} 17260--17267 ISSN 1520-6106

\bibitem{Dietrich2009}
Dietrich H~R~C, Rieger B, Wiertz F~G~M, de~Groote F~H, Heering H~A, Young I~T
  and Garini Y 2009 {\em J. Nanophotonics\/} {\bf 3} 031795

\bibitem{Uhlenbeck1930}
Uhlenbeck G~E and Ornstein L~S 1930 {\em Phys. Rev.\/} {\bf 36}(5) 823--841

\bibitem{Doob1942}
Doob J~L 1942 {\em Ann. Math.\/} {\bf 43} 351--369 ISSN 0003486X, 19398980

\bibitem{Wang1945}
Wang M~C and Uhlenbeck G~E 1945 {\em Rev. Mod. Phys.\/} {\bf 17}(2-3) 323--342

\bibitem{Barndorff2002}
Barndorff-Nielsen O~E and Shephard N 2002 {\em J. Roy. Stat. Soc. B\/} {\bf 63}
  167--241 ISSN 1369-7412

\bibitem{Maller2009}
Maller R~A, M{\"u}ller G and Szimayer A 2009 {\em Ornstein-{U}hlenbeck
  Processes and Extensions\/} (Berlin, Heidelberg: Springer Berlin Heidelberg)
  pp 421--437 ISBN 978-3-540-71297-8

\bibitem{Bartoszek2017}
Bartoszek K, Glémin S, Kaj I and Lascoux M 2017 {\em J. Theor. Biol.\/} {\bf
  429} 35--45 ISSN 0022-5193

\bibitem{Hanggi1990}
H\"anggi P, Talkner P and Borkovec M 1990 {\em Rev. Mod. Phys.\/} {\bf 62}(2)
  251--341

\bibitem{Chechkin2017}
Chechkin A~V, Seno F, Metzler R and Sokolov I~M 2017 {\em Phys. Rev. X\/} {\bf
  7}(2) 021002

\bibitem{Sposini2024PRE}
Sposini V, Nampoothiri S, Chechkin A, Orlandini E, Seno F and Baldovin F 2024
  {\em Phys. Rev. E\/} {\bf 109}(3) 034120

\bibitem{Sposini2024}
Sposini V, Nampoothiri S, Chechkin A, Orlandini E, Seno F and Baldovin F 2024
  {\em Phys. Rev. Lett.\/} {\bf 132}(11) 117101

\bibitem{Fa2003}
Fa K~S 2003 {\em Chem. Phys.\/} {\bf 287} 1--5 ISSN 0301-0104

\bibitem{Farango2014}
Farago O and Gr\o{}nbech-Jensen N 2014 {\em Phys. Rev. E\/} {\bf 89}(1) 013301

\bibitem{Uneyama2015}
Uneyama T, Miyaguchi T and Akimoto T 2015 {\em Phys. Rev. E\/} {\bf 92}(3)
  032140

\bibitem{Leibovich2019}
Leibovich N and Barkai E 2019 {\em Phys. Rev. E\/} {\bf 99}(4) 042138

\bibitem{Pacheco2022}
Pacheco-Pozo A and Sokolov I~M 2021 {\em Phys. Rev. Lett.\/} {\bf 127}(12)
  120601

\bibitem{Pacheco2023}
Pacheco-Pozo A and Sokolov I~M 2023 {\em Eur. Phys. J. B\/} {\bf 96} 152 ISSN
  1434-6036

\bibitem{MenonJr2023}
{Menon Jr} L, dos Santos M~A~F and Anteneodo C 2023 {\em J. Stat. Mech.-Theory
  E.\/} {\bf 2023} 123203

\bibitem{Pacheco2024}
Pacheco-Pozo A, Balcerek M, Wy\l{}omanska A, Burnecki K, Sokolov I~M and Krapf
  D 2024 {\em Phys. Rev. Lett.\/} {\bf 133}(6) 067102

\bibitem{Li2024}
Li M~G, Xing R, Fan L~M, Hu M, Bao J~D and Li P~C 2024 {\em New J. Phys.\/}
  {\bf 26} 093024

\bibitem{Miler2024}
Miles C~E 2024 {\em bioRxiv\/}

\bibitem{vanKampen1981}
van Kampen N~G 1981 {\em J. Stat. Phys.\/} {\bf 24} 175--187 ISSN 1572-9613

\bibitem{Tsekov1997}
Tsekov R 1997 {\em J. Chem. Soc. Faraday T.\/} {\bf 93}(9) 1751--1753

\bibitem{Sokolov2010}
Sokolov I 2010 {\em Chem. Phys.\/} {\bf 375} 359--363 ISSN 0301-0104
  {S}tochastic processes in Physics and Chemistry (in honor of {P}eter
  {H}änggi)

\bibitem{Sancho2011}
Sancho J~M 2011 {\em Phys. Rev. E\/} {\bf 84}(6) 062102

\bibitem{deHaan2012}
de~Haan H~W, Chubynsky M~V and Slater G~W 2012 Monte {C}arlo approaches for
  simulating a particle at a diffusivity interface and the
  ``{I}t{\^o}-{S}tratonovich dilemma" (\textit{Preprint}
  \eprint[http://arxiv.org/abs/1208.5081]{1208.5081})

\bibitem{Arenas2012}
Arenas Z~G and Barci D~G 2012 {\em J. Stat. Mech.-Theory E.\/} {\bf 2012}
  P12005

\bibitem{dePirey2022}
De~Pirey T~A, Cugliandolo L~F, Lecomte V and Van~Wijland F 2022 {\em Adv.
  Phys.\/} {\bf 71} 1--85

\bibitem{Escudero2023}
Escudero C and Rojas H 2023 It\^o versus {H}\"anggi-{K}limontovich
  (\textit{Preprint} \eprint[http://arxiv.org/abs/2309.03654]{2309.03654})

\bibitem{Ito1944}
It{\^o} K 1944 {\em Proc. Imp. Acad.\/} {\bf 20} 519 -- 524

\bibitem{Stratonovich1964}
Stratonovich R 1964 {\em Vestn. Mosk. U. Mat. M.\/} {\bf 1} 3--12

\bibitem{Stratonovich1966}
Stratonovich R~L 1966 {\em SIAM J. Control\/} {\bf 4} 362--371

\bibitem{Hanggi1978}
H{\"a}nggi P 1978 {\em Helv. Phys. Acta\/} {\bf 51} 183--201

\bibitem{Hanggi1980}
H{\"a}nggi P 1980 {\em Helv. Phys. Acta\/} {\bf 53} 492

\bibitem{Hanggi1982}
Hänggi P and Thomas H 1982 {\em Phys. Rep.\/} {\bf 88} 207--319 ISSN 0370-1573

\bibitem{Klimontovich1994}
Klimontovich Y~L 1994 {\em Phys.-Usp.\/} {\bf 37} 737

\bibitem{Volpe2010}
Volpe G, Helden L, Brettschneider T, Wehr J and Bechinger C 2010 {\em Phys.
  Rev. Lett.\/} {\bf 104}(17) 170602

\bibitem{Postnikov2020}
Postnikov E~B, Chechkin A and Sokolov I~M 2020 {\em New J. Phys.\/} {\bf 22}
  063046

\bibitem{Roma2014}
Rom\'a F, Cugliandolo L~F and Lozano G~S 2014 {\em Phys. Rev. E\/} {\bf 90}(2)
  023203

\bibitem{Stauffer2005}
Stauffer D and Sahimi M 2005 {\em Phys. Rev. E\/} {\bf 72}(4) 046128

\bibitem{balcerek2023modelling}
Balcerek M, Wy{\l}oma{\'n}ska A, Burnecki K, Metzler R and Krapf D 2023 {\em
  New J. Phys.\/} {\bf 25} 103031

\bibitem{Pacheco2024fBm}
Pacheco-Pozo A and Krapf D 2024 {\em Phys. Rev. E\/} {\bf 110}(1) 014105

\bibitem{Zarlenga2007}
Zarlenga D~G, Larrondo H~A, Arizmendi C~M and Family F 2007 {\em Phys. Rev.
  E\/} {\bf 75}(5) 051101

\bibitem{burov2007occupation}
Burov S and Barkai E 2007 {\em Phys. Rev. Lett.\/} {\bf 98} 250601

\bibitem{Krusemann2014}
Kr{\"u}semann H, Godec A and Metzler R 2014 {\em Phys. Rev. E\/} {\bf 89}
  040101

\bibitem{Miyaguchi2015}
Miyaguchi T and Akimoto T 2015 {\em Phys. Rev. E\/} {\bf 91}(1) 010102

\bibitem{Akimoto2016}
Akimoto T, Barkai E and Saito K 2016 {\em Phys. Rev. Lett.\/} {\bf 117}(18)
  180602

\bibitem{Akimoto2020}
Akimoto T and Saito K 2020 {\em Phys. Rev. E\/} {\bf 101}(4) 042133

\bibitem{Shafir2024}
Shafir D and Burov S 2024 {\em Phys. Rev. Lett.\/} {\bf 133}(3) 037101

\bibitem{korabel2010paradoxes}
Korabel N and Barkai E 2010 {\em Phys. Rev. Lett.\/} {\bf 104} 170603

\bibitem{burov2011time}
Burov S and Barkai E 2011 {\em Phys. Rev. Lett.\/} {\bf 106} 140602

\bibitem{Burov2017}
Burov S 2017 {\em Phys. Rev. E\/} {\bf 96}(5) 050103

\bibitem{shemer2009einstein}
Shemer Z and Barkai E 2009 {\em Phys. Rev. E\/} {\bf 80} 031108

\bibitem{Balakrishnan2008}
Balakrishnan V 2008 {\em Elements of nonequilibrium statistical mechanics\/}
  vol~3 (Springer)

\bibitem{Doi1988}
Doi M and Edwards S~F 1988 {\em The Theory of Polymer Dynamics\/} International
  series of monographs on physics (Clarendon Press) ISBN 9780198520337

\bibitem{Risken1996}
Risken H 1996 {\em The Fokker-Planck Equation: Methods of Solution and
  Applications\/} Springer Series in Synergetics (Springer Berlin Heidelberg)
  ISBN 9783540615309

\bibitem{Lidner2013}
Lindner M, Nir G, Vivante A, Young I~T and Garini Y 2013 {\em Phys. Rev. E\/}
  {\bf 87}(2) 022716

\bibitem{Lanoiselee2022}
Lanoisel\'ee Y, Stanislavsky A, Calebiro D and Weron A 2022 {\em Phys. Rev.
  E\/} {\bf 106}(6) 064127

\bibitem{balcerek2025two}
Balcerek M, Pacheco-Pozo A, Wy{\l}oma{\'n}ska A, Burnecki K and Krapf D 2025
  {\em Chaos\/} {\bf 35}

\bibitem{keyser2005nanopore}
Keyser U~F, Krapf D, Koeleman B~N, Smeets R~M, Dekker N~H and Dekker C 2005
  {\em Nano Lett.\/} {\bf 5} 2253--2256

\bibitem{Palacci2014}
Palacci J, Sacanna S, Kim S~H, Yi G~R, Pine D~J and Chaikin P~M 2014 {\em
  Philos. T. Roy. Soc. A\/} {\bf 372} 20130372

\bibitem{Vutukuri2020}
Vutukuri H~R, Lisicki M, Lauga E and Vermant J 2020 {\em Nat. Commun.\/} {\bf
  11} 2628

\bibitem{Wang2024}
Wang W, Simmchen J and Uspal W 2024 {\em Active Colloids: From Fundamentals to
  Frontiers\/} (Roy. Soc. Ch.) ISBN 978-1-83767-207-3

\bibitem{Ruemelin1982}
R\"{u}emelin W 1982 {\em SIAM J. Numer. Anal.\/} {\bf 19} 604--613

\bibitem{Garcia2012}
Garc{\'\i}a-Ojalvo J and Sancho J 2012 {\em Noise in spatially extended
  systems\/} (Springer Science \& Business Media)

\end{thebibliography}

\end{document}